\documentclass[pra,aps,twocolumn,showpacs,a4paper]{revtex4}
\usepackage{amsmath,amsfonts,amssymb,graphics,graphicx,epsfig,color,times,bbm}
\usepackage[latin1]{inputenc}

\graphicspath{{Bilder/}}

\newcommand{\ad}{\hat{b}^\dagger}
\renewcommand{\a}{\hat{b}}
\newcommand{\nb}{\hat{n}}

\newcommand{\cd}{\hat{c}^\dagger}
\renewcommand{\c}{\hat{c}}
\newcommand{\nf}{\hat{m}}

\DeclareMathSymbol{\rho}{\mathord}{letters}{"25}
\DeclareMathSymbol{\varrho}{\mathord}{letters}{"1A}

\usepackage{cancel,ifthen,ulem}
\newcommand{\Komment}[2][NoInPuT]{\ifthenelse{\equal{#1}{NoInPuT}}{}{\xcancel{\color{blue}#1}}{\color{red} #2}}

\begin{document}

\title{Fermion mediated long-range interactions of bosons in the 1D Bose-Fermi-Hubbard model}

\author{Alexander Mering}
\affiliation{Fachbereich Physik and research center OPTIMAS, Technische Universit\"at Kaiserslautern, D-67663 Kaiserslautern, Germany}
\email{amering@physik.uni-kl.de}
\author{Michael Fleischhauer}
\affiliation{Fachbereich Physik and research center OPTIMAS, Technische Universit\"at Kaiserslautern, D-67663 Kaiserslautern, Germany}

\begin{abstract}
The ground-state phase diagram of mixtures of spin polarized fermions and bosons in a 1D periodic lattice is discussed in the limit of large fermion hopping and half filling of the fermions. Numerical simulations performed with the density matrix renormalization group (DMRG) show besides  bosonic Mott insulating (MI), superfluid (SF), and charge density-wave phases (CDW) a novel phase with spatial separation of MI and CDW regions. 
We derive an effective bosonic theory which allows for a complete understanding and
quantitative prediction of the bosonic phase diagram. In particular the origin of 
CDW phase and the MI-CDW phase separation is revealed as the interplay between 
fermion-induced mean-field potential and long range interaction with alternating sign. 
\end{abstract}


\date{\today}

\maketitle

Ultra-cold atomic gases in light induced periodic potentials have become an important experimental testing ground
for concepts of many-body physics since they allow the realization of precisely controllable 
model Hamiltonians with widely tunable parameters. A system which has attracted particular interest in the recent past is a mixture of bosons and spin-polarized fermions in a deep lattice potential, described by the Bose-Fermi-Hubbard model (BFHM) \cite{Albus2003,Lewenstein2004}. Mixing lattice bosons with fermions, G\"unther {\it et al.} \cite{Guenther2006} and Ospelkaus {\it et al.} \cite{Ospelkaus2006} observed an unexpected reduction of bosonic superfluidity which triggered a number of theoretical and experimental studies on the influence of fermions on boson superfluidity.  In the limit of small fermion mobility the phase diagram
can be well understood by mapping to the purely bosonic Hubbard system with binary disorder \cite{Mering2008,Krutitsky2008}.
For increasing fermionic hopping amplitudes a number of new phenomena emerge  \cite{Pollet2006}, including polaronic
phases \cite{Mathey2004} and density waves \cite{Pazy2005, Hebert2007}. Furthermore under certain conditions 
bosons can enter the a supersolid phase (SS), where CDW and off-diagonal long-range 
order coexist \cite{Buechler2003,Titvinidze2008,Hebert2008}. \\
In the present paper we study the 1D BFHM  in the limit of large fermion hopping which allows for a rather 
comprehensive understanding of the existing phases and their origin in particular in the case of half filling of the spin-polarized fermions. 
In the large hopping limit the fermions can be formally integrated out \cite{Buechler2003,Buechler2004,Lutchyn2008} resulting in an oscillating mean-field potential as well as a
long-range density-density interaction between the bosons. This interaction has  
alternating sign if the fermion filling is commensurate with the lattice which is the origin of the 4$k_F$ CDW. 
In the thermodynamic limit it is however formally divergent and  needs to be renormalized which is done here by taking into account the back-action of the bosons to the fast fermions. The resulting effective boson model allows an analytic and quantitative
prediction of the $(\mu_B-J_B)$ phase diagram, where $\mu_B$ is the bosonic chemical potential and
$J_B$ the corresponding hopping amplitude. At double-half filling, i.e. $\rho_F=\rho_B=\frac 12$, we identify an incompressible
CDW phase and study its transition to a SF with increasing $J_B$ both using analytic results
from the effective model and numerical simulations based on DMRG \cite{Schollwoeck2005}. 
DMRG simulations also
show the presence of a novel phase with coexistence between spatially separated Mott-insulator and  CDW regions for non-commensurate boson filling.  This phase which is absent for an extended Bose-Hubbard model
\cite{Kuehner2000} can be well explained within the effective model and is shown to exist for all values of the boson-fermion interaction.
As the effective theory describes the appearance of a density wave on a quantitative level it is expected to
explain also the conditions for the existence of a SS found in \cite{Hebert2008} using quantum Monte Carlo simulations. 
Numerical evidence was given \cite{Hebert2008} that a SS  exists only if the filling of spin-polarized fermions 
is exactly one half but that of bosons is not commensurate and if the boson-fermion repulsion exceeds that of the 
bosons. The emergence of a finite superfluid fraction from the effective model will be discussed 
at a different place \cite{MeringUnpublished}.

Mixtures of ultracold bosons and spin-polarized fermions in optical lattices are well described by the Bose-Fermi-Hubbard Hamiltonian \cite{Jaksch1998,Bloch2008}
\begin{align}\label{eq:BFHM}
\hat H&=-J_B\sum_j\left(\ad_j\a_{j+1}+\ad_{j+1}\a_{j}\right)+\frac{U}{2}\sum_j\nb_j\left(\nb_j-1\right)\nonumber\\
	&\hspace{0.5cm}-J_F\sum_j\left(\cd_j\c_{j+1}+ \cd_{j+1}\c_{j}\right)+V\sum_j\nb_j\nf_j,
\end{align}
where $\ad,\a$ ($\cd,\c$) are bosonic (fermionic) creation and annihilation operators and $\nb$ ($\nf$) the corresponding number operators. Here, the bosonic (fermionic) 
hopping amplitude is given by $J_B$ ($J_F$), and $U$ ($V$) accounts for the intra- (inter-) species interaction energy. In the following
we consider the limit of large fermionic hopping, i.e. we assume $J_F\gg U,|V|,J_B$ and the energy scale is set by $U=1$. \\
  \begin{figure}[ht]
   \epsfig{file=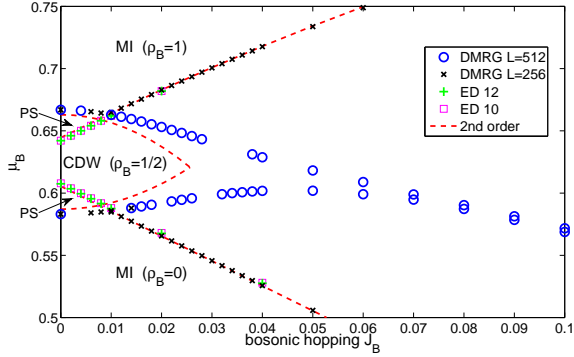,width=1\columnwidth}
\caption{(Color online) Boundaries of the incompressible MI phases and the CDW phase for half fermion filling $\rho_F=1/2$, 
$J_F=10$, and $V=1.25$ obtained by DMRG and for small values of $J_B$ by ED.
One recognizes partial overlap between MI and CDW phase for small values of $J_B$
indicating regions of spatial phase separation (PS) between MI and CDW. The dashed lines are results from the 2nd order perturbation theory based on the effective bosonic model}
\label{fig:IncompressiblePhase}
 \end{figure}

\begin{figure}[ht]
   \epsfig{file=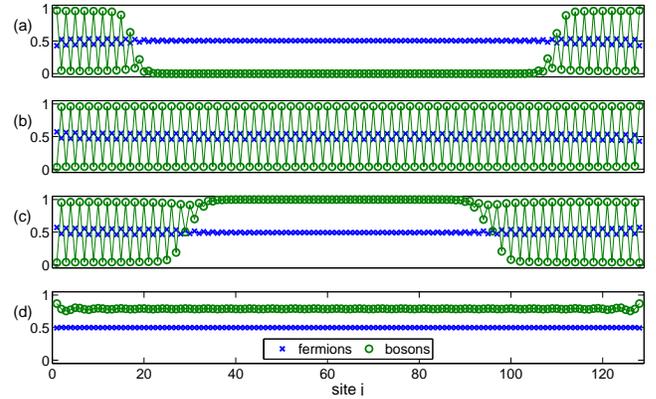,width=1\columnwidth}
 \caption{(Color online) Densities of bosons and fermions corresponding to the different regions in the boson phase 
diagram fig. \ref{fig:IncompressiblePhase} for a lattice of length $L=128$ and open boundary conditions. The boson number $N_B$ is (a) 19, (b) 64, (c) 96, and (d) 101. The plots (a,b,c) are taken at $J_B=0.01$ and (d) at $J_B=0.07$. While (b) displays the gaped CDW, (a) and (c) shows the PS phase. (d) is outside of the parameter regime with a CDW.} 
\label{fig:DensityCuts}
 \end{figure}

In this limit of large fermionic hopping the physics of the BFHM is well captured by the bosonic phase diagram alone. Considering the most
interesting case of half filling of the spin-polarized fermions, we have plotted 
in figure \ref{fig:IncompressiblePhase} the phase diagram for the bosons obtained numerically by DMRG simulations and exact diagonalization (ED)
for $J_F=10$, $V=1.25$. Besides the MI and SF phases expected from the pure bosonic model, the phase diagram displays a CDW phase at double-half filling ($\rho_F=\rho_B=1/2$).  
Exact diagonalization is used for very small $J_B$ to avoid boundary effects. Furthermore a novel phase is visible for the case of non-commensurate boson filling in which spatially separated regions
of bosonic Mott insulators and density waves coexist (phase separation, PS). Figure \ref{fig:DensityCuts}  shows numerical results for the density cuts from within the corresponding phases. We note that while the pinning of the CDW to the boundaries is a result of the open boundary conditions required for DMRG, the phase separation persits for large systems and was verified for small systems using periodic boundary conditions.   \\

In the following we derive an effective bosonic model which provides an understanding of the phase diagram in figure \ref{fig:IncompressiblePhase} on a quantitative level. In the limit of fast 
fermions one could expect that their main influence is through a mean field contribution, which according to eq. (\ref{eq:BFHM}) amounts to a simple shift of the bosonic chemical potential $\mu_B \to \mu_B -\rho_F\ V$. Indeed the two Mott lobes in fig. \ref{fig:IncompressiblePhase}
are symmetrically located around $\mu = \frac{1}{2}\ V $. To explain the CDW and PS phases one needs however an
effective description beyond the mean-field level. In order to adiabatically eliminate the fast fermions we first separate
the fermion Hamiltonian. As will be seen later on it is essential to take into account the back action of bosons to
the fermions. To do this we incorporate in the fermion Hamiltonian the interaction with a mean field
potential given by a yet undetermined density $\widetilde{n}_j$ of bosons. Thus $\hat H_{\rm F} =-J_F\sum_j\left(\cd_j\c_{j+1}+ \cd_{j+1}\c_{j}\right)+V\sum_j \widetilde n_j\nf_j$ is the fermionic Hamiltonian and $\hat H_{\rm I} =V\sum_j \left(\nb_j-\widetilde n_j\right)\nf_j$ the interaction.
We start
from the S-matrix $\hat{\mathcal S} ={ \mathcal T}\exp\{-\frac i\hbar \int_{-\infty}^\infty {\rm d}\tau  H_I(\tau)\}$  of the full system with  $\hat H_{\rm I}(\tau) = e^{-\frac{i}{\hbar}(\hat H-\hat H_{\rm I})\tau} \hat H_{\rm I} e^{\frac{i}{\hbar} (\hat H-\hat H_{\rm I})\tau}$ and trace out the fermions exactly using a cumulant expansion \cite{Gardiner1985}.  Since the characteristic time scale of fermionic correlations is of order $1/J_F$ and thus much shorter than any other time scale in the system, a Markov approximation can be used, replacing two-time bosonic 
operators by equal time operators, i. e.
$ \int {\rm d} \tau \int {\rm d}\tau^\prime\  \hat n_j(\tau) \hat n_{j+d}(\tau^\prime) \left\langle\left\langle \hat m_j(\tau) \hat m_{j+d}(\tau^\prime) \right\rangle\right\rangle_{\rm F}$\\ $
{ \mapsto} \int {\rm d} \tau\   \hat n_j(\tau) \hat n_{j+d}(\tau)\   \int {\rm d}\tau^\prime\left\langle\left\langle \hat m_j(\tau) \hat m_{j+d}(\tau^\prime) \right\rangle\right\rangle_{\rm F}$.
This leads to an effective Hamiltonian for the bosons 
\begin{align}
  \hat{H}_{\rm B}^{\rm eff}&=\hat H_{\rm B} + V\sum_j \left(\nb_j-\widetilde n_j\right)\left\langle 
\nf_j\right\rangle_{\rm F}\label{eq:effectiveBosonic}\\
  &\hspace{1cm} +\sum_j\sum^\infty_{l=-\infty} g_l \left(\nb_j-\widetilde n_j\right)\left(\nb_{j+l}-\widetilde 
n_{j+l}\right),\notag
 \end{align}
 where $ g_l = - i \frac{V^2}{2\hbar}\int_{-\infty}^\infty {\rm d}\tau \left<\left<{\cal T} 
\nf_j(\tau)\nf_{j+l}(0)\right>\right>_{\rm F} $.
${\cal T}$ denotes time-ordering. One recognizes a fermion-induced mean-field potential proportional to 
$\langle \hat m_j\rangle_F$.
The couplings $g_l$ describes a long-range density-density interaction between the bosons separated by $l$ lattice sites. The mean-field potential and the 
density-density interaction are the only interaction terms emerging in the effective theory since higher order moments in the cumulant expansion vanish. \\

In the case of free fermions, i.e. ignoring the back-action of bosons ($\widetilde n_j \equiv 0$),  fermionic correlations and couplings $g_l$ can easily be calculated. For $\rho_F=1/2$, $g_l$ scales asymptotically as $g_l\sim {(-1)^l}/l$, i.e. has a long-range character and alternating sign. 
For a general fermion density $\rho_F$ they oscillate 
with period $1/\rho_F$ which is typical for induced interactions of the RKKY type \cite{Rudermann1954,Kasuya1956,Yosida1957}. The oscillation of the interaction is the origin of the formation of charge
density waves. For $\rho_F=1/2$ the interaction energy is minimized if the bosons occupy sites with
distance $2$. An effective theory with coupling constants resulting from
free fermions has however a fundamental problem:
As $g_l\sim 1/l$ the boson-boson interaction energy diverges 
logarithmically with the total length of the lattice.
This would result in an incompressible CDW for {\it any} value of the bosonic hopping $J_B$. Thus such a theory completely fails to describe the
transition from a CDW phase to a bosonic superfluid. An accurate description of this transition is however
important e.g. to explain the PS phase and to address existence conditions of a supersolid phase.
Therefore it is necessary to renormalize the effective
interaction. This is done here by taking into account the back action of the bosons through the mean-field potential in $\hat H_F$. Since for $\rho_F=1/2$ the bosonic system is driven 
into a CDW with period 2, a good ansatz is $\widetilde n_j = \frac12\bigl[1 +  (-1)^j\eta \bigr]$, 
where $\eta=\eta(J_B)$ is the amplitude of the density oscillation with $\eta\bigr|_{J_B=0}=1$. In general $\eta$ is treated as a free parameter
and can be determined self-consistently by a minimization of the energy. For the following
calculations it turns out to be more convenient to introduce a parameter $a$ proportional to $\eta$ as $a={\eta \, V}/{(4\sqrt{2\pi}J_F)}$.  \\

In order to calculate the fermionic cumulants and therefore $g_l$ analytically from $H_{\rm f}$ in the presence of the boson-induced mean field potential we make use of Green's function techniques and perturbation theory in $\eta V/{J_F}$ (for details see \cite{MeringUnpublished}). This gives $g_l$ as a function of $a$ as
 \begin{multline}
    g_l(a) = -\frac{V^2}{8\pi^2  J_F }\times\\
    \int_{0}^{\pi}\int_{0}^{\pi}{\rm d}\xi{\rm d}\xi^\prime\,  
\frac{\cos(\xi l)\,\cos(\xi^\prime l)}{ \sqrt{\cos^2(\xi)+a^2}+\sqrt{\cos^2(\xi^\prime)+a^2}}\times\\
\left(1+ \frac{\cos(\xi)}{\sqrt{\cos^2( \xi)+a^2}}\right)\left(1-\frac{\cos(\xi^\prime)}{\sqrt{\cos^2( \xi^\prime)+a^2}}\right).\label{eq:couplingsrenormalized}
 \end{multline}
Similarly one finds for the fermionic density
%
%
\begin{equation}
 \left\langle \nf_j\right\rangle =\frac12\Bigl[1-(-1)^j \eta_F\Bigr],\label{eq:f-density}
\end{equation}
 with $\eta_F =\frac{4a}{\pi\sqrt{1+a^2}}K\Bigl[\frac{1}{1+a^2}\Bigr]$ and $K[x]$ being the complete elliptic integral of the first kind. This equation along with
\begin{eqnarray}
\left\langle \nb_j\right\rangle = \frac12\Bigl[1+(-1)^j \eta\Bigr] = \widetilde n_j\label{eq:b-density}
\end{eqnarray}
gives the density distributions of fermions and bosons for double-half filling, i.e. in
the CDW phase, as function of the variational parameter $a$.
In the limit $a \to 0$ the above expressions reduces to the free fermion case ($\left\langle \nf_j\right\rangle=\frac12$).
 \begin{figure}[ht]
   \epsfig{file=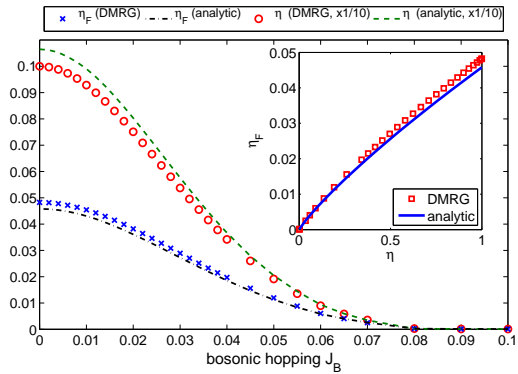,width=1\columnwidth}
  \caption{(Color online) CDW amplitudes (points) of bosons and fermions for
double half filling as function of normalized boson hopping. Dash-dotted line: calculated factor $a$ from the fermionic data. Dashed line: analytic estimate of the bosonic amplitude from the fermionic input $a$. One recognizes a very good agreement between the numerical data and the analytic estimate, where it should be kept in mind that the underlying perturbation theory gets better for $\eta\to0$, i.e. for {\it increasing} $J_B $.\label{fig:Amplitude}}
 \end{figure}
To provide a first test of the validity of the effective theory we have plotted in Fig. \ref{fig:Amplitude} 
the ratio of the amplitudes of the bosonic and fermionic density waves obtained from the data in fig. \ref{fig:IncompressiblePhase} along with the
prediction from eqs. (\ref{eq:f-density}) and (\ref{eq:b-density}). Note that this ratio is exactly fixed by the effective theory and independent on the variational parameter $a$. One recognizes a very good agreement. Also shown in Fig. \ref{fig:Amplitude} are the 
amplitudes of bosonic and fermionic density waves respectively as function of bosonic hopping $J_B$ obtained numerically
as well as from the effective theory using the numerical data of the other species as input.  It can be seen, that the renormalized effective theory fits quite well with the numerical results.\\

Using the effective Hamiltonian, (\ref{eq:effectiveBosonic}) we will now derive an analytic approximation to the phase diagram of the full BFHM using a  strong-coupling 
expansion valid  for small values of $J_B$ \cite{Freericks1996}. Since we are mainly interested in the boundaries of the incompressible lobes, we will calculate them in the canonical ensemble from the energies of the relevant states as a function of the bosonic hopping amplitude $J_B$. The upper (lower) boundary is given by the bosonic particle-hole gap. With this, the chemical potentials for bosonic filling $\rho_B$ are given by $ \mu_{\rho_B}^\pm = \pm \left( E(\rho_B L\pm 1)-E(\rho_B L)\right)$, where $E(N)$ is the ground state energy for a given number of bosons $N$. At $J_B=0$, this is straight forward to calculate since the ground state distribution of the bosons is trivial and the variation parameter $\eta$ is fixed to unity.

For $J_B>0$ we apply degenerate perturbation theory in 2nd order using Kato's expansion \cite{Klein1974}. In second order, there is a local correction of the ground state energy for all numbers of particles, as well as 2nd order two-site hopping processes connecting the states within the ground state manifold in the case of an additional (absent) boson. Incorporating this, the upper and lower chemical potentials for the CDW phase can be expressed in a simple analytic form as
\begin{eqnarray}
  \mu^\pm_\frac{1}{2}=\frac V2 \pm V\, \eta_F \pm g_0(a) - \beta_\pm J_B^2.\label{eq:mu-1/2}
\end{eqnarray}
Similarly one finds for the chemical potential corresponding to unity and zero filling 
\begin{eqnarray}
 \mu_{1}^- &=& \frac V2 - g_0(0) + 2 J_B - \alpha J_B^2,\\
\mu_{0}^+&=& \frac V2 + g_0(0) - 2 J_B.\label{eq:mu+0}
\end{eqnarray}
Here $\eta_F$ and $g_0(a)$ are taken for $\eta\equiv 1$ $(J_B=0)$ because of the perturbation expansion. The derivation of $\alpha$ and $\beta_\pm$ is lengthy but straight forward.
Their explicit form will be given along with other details elsewhere \cite{MeringUnpublished}. One finds in particular $\beta_+>0>\beta_-$. Note that $V\eta_F$ is positive irrespective of the sign of $V$ and is larger in magnitude than both $g_0(0)$ and $g_0(a)$. With this one can see that $ \mu^+_\frac{1}{2}\bigr|_{J_B=0}>\mu_1^-\bigr|_{J_B=0}$, $ \mu^-_\frac{1}{2}\bigr|_{J_B=0}<\mu_0^+\bigr|_{J_B=0}$.
Thus there exists a region where the chemical potential is not monotonous in the boson number. This explains the coexistence of MI and CDW in the PS phases found numerically in fig. \ref{fig:IncompressiblePhase}.  
The long-range character of the fermion mediated interaction together with the fermion-induced mean-field potential prefers extended, spatially homogeneous regions
of a commensurate CDW. Extra bosons will be pushed out and form an incompressible Mott insulator region. Such a phase does not exist in purely bosonic systems with extended interactions due to the
absence of the fermion induced mean-field potential $V\, \eta_F$.
If the bosonic hopping exceeds a certain critical value, given by the crossing of the curves $\mu^\pm_\frac{1}{2}$
with $\mu_1^-$ or respectively $\mu_0^+$ the minimization of kinetic energy by equally distributing the
particle is larger than the loss in interaction energy due to the fermion mediated interaction. \\
  \begin{figure}
 \epsfig{file=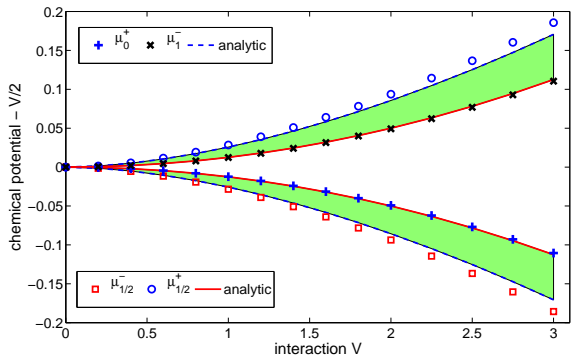,width=\columnwidth}
\caption{(Color online) Boundaries (shifted by $V/2$) of incompressible MI and CDW phases for half fermion filling $\rho_F=1/2$ at vanishing bosonic hopping $J_B=0$ with varying interaction $V$. Curves are the analytic results and the data points are obtained by DMRG (CDW phase, $L=128$) and ED (MI lobes, periodic boundaries with infinite size scaling). The shaded region depicts the coexistence phase between MI and CDW with $\mu^+_0 > \mu^-_\frac12$ and $\mu^+_\frac12 > \mu^-_1$.}
\label{fig:Chemicalpotentials}
 \end{figure}

In fig. \ref{fig:Chemicalpotentials} we have plotted the
chemical potentials for zero bosonic hopping as function of the interaction strength $V$ defining the boundaries of the Mott insulating phases with 
zero and unity filling as well as the lower and upper boundaries of the CDW phase with half filling of bosons. 
One recognizes that phase separation between MI and CDW exists for all values of the boson-fermion
interaction $V$. Once again there is a rather good agreement between full numerics and effective theory,
which provides another test for its validity.

The phase boundaries for $J_B>0$ obtained from the analytic results for the chemical potentials in eqns. (\ref{eq:mu-1/2}) to (\ref{eq:mu+0}) are shown in fig. \ref{fig:IncompressiblePhase} as dashed lines. Although the precise form of the CDW lobe is not correctly reproduced (as expected for the strong-coupling perturbation approach), the qualitative agreement is remarkable.  The strong coupling approximation yields a critical value of  $J_B^{\rm  CDW}=0.025$
beyond which the CDW ceases to exist  for $J_F=10$ and $V=1.25$. Whether or not the CDW gap vanishes at a finite value of $J_B$ is however unclear. Our numerics indicates that a very small gap may persist even for values of $J_B=1$ and may not
close at for $J_B<J_F$.
The critical values $J_B^{\rm PS}\approx 0.01$ for the PS region obtained from the effective model agrees however rather well
with the numerical data.\\

In summary we developed an effective model for a mixture of bosons and spin polarized fermions 
in a periodic lattice in the limit of large fermion hopping. This model reveals the physical
origin of the incompressible CDW phase and provides a simple quantitative description. The fast fermions 
cause a mean-field potential and mediate a long-range density-density interaction which is  of alternating sign for $\rho_F=1/2$. 
In order to accurately describe the conditions for the existence of a  CDW  renormalization
effects due to the back-action of the bosons need to be taken into account.
The density wave amplitudes where obtained
from an analytic model and verified by numerical DMRG simulations. The effective model also gives 
a simple understanding and quantitative description of a novel phase where spatially separated regions
of a maximum amplitude CDW and a MI coexist. The effective model is expected to provide a
means for predicting and understanding conditions for the existence of a SS phase in Bose-Fermi mixtures and other mass imbalanced two-species models.

\section*{Acknowledgements}

This work has been supported by the DFG through the SFB-TR 49 and the GRK 792. 
We also acknowledge the computational support from the NIC at FZ J{\"u}lich and thank 
U. Schollw\"ock for his DMRG code. Furthermore we thank E. Altmann, W. Hofstetter, C. Kollath, M. Snoek and T. Giamarchi for useful discussions.

\bibliographystyle{apsrev}

\end{document}